\documentclass[epj]{webofc} 
\usepackage[varg]{txfonts}   % Web of Conferences font 

\woctitle{HYP2022 - 14$^\textrm{th}$ International Conference on Hypernuclear
and Strange Particle Physics}

\def\lamb#1#2{$^{#1}_{\Lambda}${#2}}
\def\lam#1#2{$^{#1}_{~\Lambda}${#2}}

\usepackage{slashed}

\begin{document} 

\title{$\Xi$-nuclear constraints from $\Xi^-$ emulsion capture events} 
\author{\firstname{Eliahu} \lastname{Friedman}\inst{1}\fnsep
\thanks{Eliahu.Friedman@mail.huji.ac.il} \and \firstname{Avraham} 
\lastname{Gal}\inst{1}\fnsep\thanks{avragal@savion.huji.ac.il} 
\institute{Racah Institute of Physics, The Hebrew University, 
Jerusalem 9190401, Israel}} 

\abstract
{All five KEK and J-PARC two-body $\Xi^-$+$^A$Z $\to$ \lam{A'}{Z'}+
\lam{A''}{Z''} capture events in light emulsion nuclei, including KISO and 
IBUKI in $^{14}$N, are consistent with Coulomb-assisted $1p_{\Xi^-}$ nuclear 
states. The underlying $\Xi$-nuclear potential is strongly attractive, with 
nuclear-matter depth $V_{\Xi}$ larger than 20 MeV. The recent $^{14}$N capture 
events KINKA and IRRAWADDY assigned by J-PARC E07 to $1s_{\Xi^-}$ nuclear 
states, and implying considerably shallower $V_{\Xi}$, have also another 
interpretation as $1p_{\Xi^0}$ nuclear states.} 
\maketitle

\section{Introduction: $\Xi^-$ capture events} 
\label{sec:intro} 

The nuclear interactions of $\Xi$ hyperons are poorly known~\cite{GHM16,HN18}. 
Because of the large momentum transfer in the standard $(K^-,K^+)$ production 
reaction, induced by the two-body $K^-p\to K^ + \Xi^-$ strangeness exchange 
reaction, the $\Xi^-$ hyperons are produced dominantly in the quasi-free 
continuum region, with less than 1\% expected to form $\Xi^-$-nuclear bound 
states that decay by a $\Xi^- p\to\Lambda\Lambda$ strong-interaction capture 
reaction. In fact, no $\Xi^-$ or $\Lambda\Lambda$ hypernuclear bound states 
have ever been observed unambiguously in such experiments, e.g., KEK-PS 
E224~\cite{E224} and BNL-AGS E885~\cite{E885}, both on $^{12}$C, and BNL-AGS 
E906~\cite{E906} on $^9$Be. A potential depth of $V_{\Xi}=17\pm 6$~MeV was 
deduced recently from the quasi-free $\Xi^-$ spectrum taken in the 
$^9$Be($K^-,K^+$) reaction~\cite{HH21}. 

Here we focus on $\Xi$ nuclear constraints derived by observing $\Xi^-$ 
capture events in exposures of light-emulsion CNO nuclei to the $(K^-,K^+)$ 
reaction. A small fraction of the produced high-energy $\Xi^-$ hyperons 
slows down in the emulsion, undergoing an Auger process to form high-$n$ 
atomic states, and cascades down radiatively. Strong-interaction capture 
takes over atomic radiative cascade in a 3D atomic orbit bound by 126, 175, 
231~keV in C, N, O, respectively, affected to less than 1~keV by the strong 
interaction~\cite{BFG99}. Interestingly, all two-body $\Xi^-$ capture events 
$\Xi^-+{^{A}Z}\to$~\lam{A'}{Z'}+\lam{A''}{Z''} to twin single-$\Lambda$ 
hypernuclei reported in KEK and J-PARC light-emulsion $K^-$ exposure 
experiments~\cite{E176,E373a,E373b,E07a,E07b} are consistent with $\Xi^-$ 
capture from a lower orbit: a Coulomb-assisted $1p_{\Xi^-}$ nuclear state 
bound by $\sim$1~MeV. These capture events are listed in Table~\ref{tab:twin}. 
Expecting the final two $\Lambda$ hyperons in $\Xi^-p\to \Lambda\Lambda$ 
capture to form in a spin $S=0,\,1s_{\Lambda}^2$ configuration, the initial 
$\Xi^-$ hyperon and the proton on which it is captured must satisfy 
$l_{\Xi^-}=l_p$~\cite{Zhu91}, which for $p$-shell nuclear targets favors the 
choice $l_{\Xi^-}=1$. Not listed in the table are multi-body capture events 
that require for their interpretation undetected capture products, usually 
neutrons, on top of a pair of single-$\Lambda$ hypernuclei. Several of these 
new J-PARC E07 events~\cite{E07b}, like KINKA and IRRAWADDY, imply $\Xi^-$ 
capture from $1s_{\Xi^-}$ nuclear states, with estimated capture rates of 
order 1\% of capture rates from the $1p_{\Xi^-}$ nuclear states considered 
here~\cite{Zhu91,Koike17}. We discuss these states below. 

\begin{table}[!h] 
\begin{center}
\caption{Two-body $\Xi^-$ capture emulsion events from KEK and J-PARC 
experiments.} 
\begin{tabular}{ccccc} 
\hline
Experiment & Event & $^{A}Z$ & \lamb{A'}{Z'}+\lamb{A''}{Z''} & $B_{\Xi^-}$
(MeV)  \\
\hline
KEK E176~\cite{E176} & 10-09-06 & $^{12}$C & \lamb{4}{H}+\lamb{9}{Be} 
& 0.82$\pm$0.17  \\
KEK E176~\cite{E176} & 13-11-14 & $^{12}$C & \lamb{4}{H}+\lamb{9}{Be}$^{\ast}$ 
& 0.82$\pm$0.14  \\
KEK E176~\cite{E176} & 14-03-35 & $^{14}$N & \lamb{3}{H}+\lam{12}{B} 
& 1.18$\pm$0.22  \\
KEK E373~\cite{E373b} & KISO & $^{14}$N & \lamb{5}{He}+\lam{10}{Be}$^{\ast}$ 
& 1.03$\pm$0.18  \\
J-PARC E07~\cite{E07a} & IBUKI & $^{14}$N & \lamb{5}{He}+\lam{10}{Be} 
& 1.27$\pm$0.21  \\
\hline 
\end{tabular} 
\label{tab:twin} 
\end{center} 
\end{table}

\section{$\Xi$ nuclear optical potential} 
\label{sec:Vopt} 
 
$\Xi^-$ atomic and nuclear bound states are calculated using a standard 
$t\rho$ optical-potential form~\cite{FG21} 
\begin{equation} 
V_{\rm opt}(r)=-\frac{2\pi}{\mu}\,(1+\frac{A-1}{A}\frac{\mu}{m_N})\,
[b_0\,\rho(r)+b_1\,\rho_{\rm exc}(r)], 
\label{eq:Vopt} 
\end{equation}
where $\mu$ is the $\Xi^-$-nucleus reduced mass and the complex strength 
parameters $b_0$ and $b_1$ are effective, generally density dependent 
$\Xi N$ isoscalar and isovector c.m. scattering amplitudes respectively. 
The density $\rho=\rho_n+\rho_p$ is a nuclear density distribution 
normalized to the number of nucleons $A$ and $\rho_{\rm exc}=\rho_n-\rho_p$ 
is a neutron-excess density with $\rho_n=(N/Z)\rho_p$, implying that 
$\rho_{\rm exc}=0$ for the $N=Z$ emulsion nuclei $^{12}$C and $^{14}$N 
considered here. A finite-size Coulomb potential $V_c$, including 
vacuum-polarization terms is added. For densities we used mostly 
harmonic-oscillator (HO) densities~\cite{Elton61} where the r.m.s. radius 
of $\rho_p$ was set equal to that of the nuclear charge density~\cite{AM13}. 
Folding reasonably chosen $\Xi N$ interaction ranges other than corresponding 
to the proton charge radius, or using Modified Harmonic Oscillator densities, 
or replacing HO densities by realistic three-parameter Fermi density 
distributions, made little difference: all the calculated binding energies 
changed by a small fraction, about 0.03~MeV, of the uncertainty imposed by 
the $\pm$0.15 MeV experimental uncertainty of the 0.82~MeV $1p_{\Xi^-}$ 
binding energy in $^{12}$C listed in Table~\ref{tab:twin}. This holds also 
for adding a $\rho_{\rm exc}\neq 0$ term induced by considering realistic 
differences of neutron and proton r.m.s. radii.

Accepting the binding energy interval $B_{\Xi^-}^{1p}$=0.82$\pm$0.15~MeV for 
the two KEK E176 events listed in Table~\ref{tab:twin}, a $\Xi$-nuclear 
potential strength of Re$\,b_0=0.32\pm 0.01$~fm follows for a fixed value 
Im$\,b_0=0.01$~fm. The sensitivity to variations of Im$\,b_0$ is minimal: 
choosing Im$\,b_0=0.04$~fm~\cite{BFG99} instead of 0.01~fm increases Re$\,b_0$ 
by 0.01~fm to $0.33\pm 0.01$~fm. The value Re$\,b_0=0.32\pm 0.01$~fm implies 
in the limit $A\to\infty$ and $\rho(r)\to\rho_0$=0.17~fm$^{-3}$ a depth value 
$V_{\Xi}=24.3\pm 0.8$~MeV in nuclear matter, compatible with that derived 
from AGS-E906 in Ref.~\cite{HH21} and in agreement with the range of values 
21--24~MeV extracted from old emulsion events~\cite{DG83}. 

So far we have discussed a density independent $t$-matrix element $b_0$ in 
$V_{\rm opt}$, Eq.~(\ref{eq:Vopt}), to fit the $\Xi^-$ capture events in 
$^{12}$C from Table~\ref{tab:twin}. To explore how robust the deduced $\Xi$ 
potential-depth value $V_{\Xi}=24.3\pm 0.8$~MeV is, we introduce the next to 
leading-order density dependence of $V_{\rm opt}$, replacing Re$\,b_0$ 
in Eq.~(\ref{eq:Vopt}) by
\begin{equation} 
{\rm Re}\,b_0(\rho)=\frac{{\rm Re}\,b_0}
{1+\frac{3k_F}{2\pi}{\rm Re}\,b_0^{\rm lab}},\,\,\,\,\,\,\,\,\,\,
k_F=(3{\pi}^2\rho/2)^{\frac{1}{3}}, 
\label{eq:WRW} 
\end{equation}
where $k_F$ is the Fermi momentum corresponding to nuclear density $\rho$ 
and $b_0^{\rm lab}=(1+\frac{m_{\Xi^-}}{m_N})b_0$ is the lab transformed 
form of the c.m. scattering amplitude $b_0$. Eq.~(\ref{eq:WRW}) 
accounts for Pauli exclusion correlations in $\Xi N$ in-medium multiple 
scatterings~\cite{DHL71,WRW97}. Shorter-range correlations, disregarded here, 
were shown in Ref.~\cite{WH08} to contribute less than $\sim$30\% of the 
long-range Pauli correlation term. Applying Eq.~(\ref{eq:WRW}) in the present 
context, $B_{\Xi^-}^{1p}(^{12}$C)=0.82~MeV is refitted by Re$\,b_0$=0.527~fm. 
The nuclear-matter $\Xi$-nuclear potential depth $V_{\Xi}$ decreases from 
24.3$\pm$0.8 to 21.9$\pm$0.7~MeV, a decrease of merely 10\%.

\section{$1p_{\Xi^-}$ states in $^{14}$N} 
\label{sec:14N} 

\begin{figure}[!h] 
\begin{center} 
\includegraphics[width=0.5\textwidth]{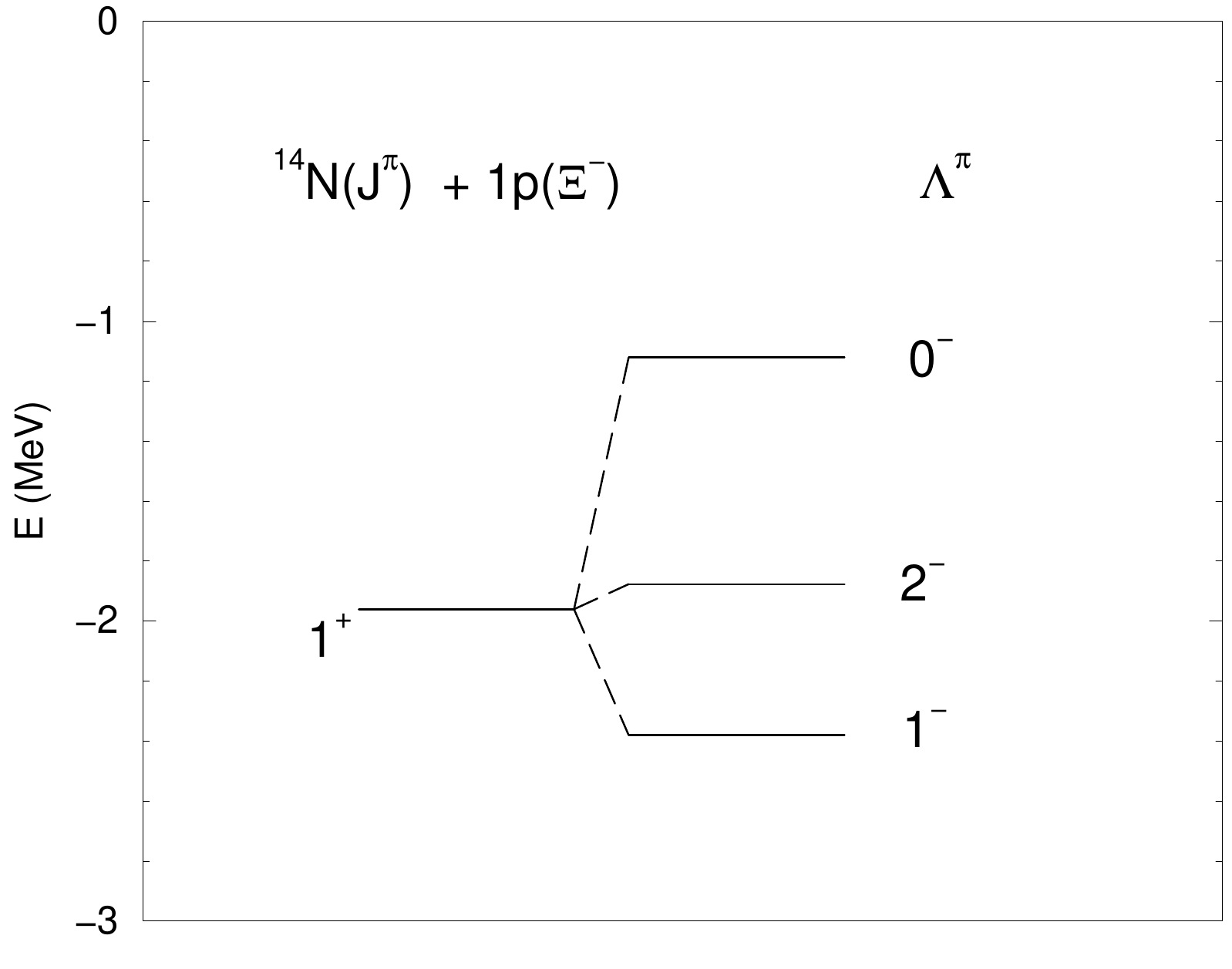} 
\caption{Energies (in MeV) of $\Lambda^{\pi}=(0^-,1^-,2^-)$ triplet of
$^{14}$N$_{\rm g.s.} + 1p_{\Xi^-}$ states, split by a $Q_N\cdot Q_{\Xi}$
residual interaction (\ref{eq:QdotQ}). The $(2\Lambda+1)$-averaged energy
$-1.96$~MeV was calculated using the same Pauli-corrected optical potential
parameter $b_0$ that yields a $^{12}$C$_{\rm g.s.} + 1p_{\Xi^-}$ state at
$-$0.82~MeV, corresponding to the $\Xi^-$ capture events in $^{12}$C listed 
in Table~\ref{tab:twin}. Figure updating Fig.~2 in Ref.~\cite{FG21}.}
\label{fig:14N}
\end{center}
\end{figure}

Applying Eqs.~(\ref{eq:Vopt},\ref{eq:WRW}) to $^{14}$N, with Re$\,b_0$ fitted 
to $B_{\Xi^-}^{1p}(^{12}$C)=0.82$\pm$0.15~MeV, results in $B_{\Xi^-}^{1p}
(^{14}$N)=1.96$\pm$0.26~MeV, Pauli correlations included. This binding energy 
is considerably higher than the value $B_{\Xi^-}$=1.15$\pm$0.20~MeV obtained 
from the three events assigned in Table~\ref{tab:twin} to $\Xi^-$ capture in 
$^{14}$N. To resolve this apparent dicrepancy, we note that the calculated 
$B_{\Xi^-}^{1p}(^{14}$N) corresponds to a $(2\Lambda+1)$-average of binding 
energies for a triplet of states $\Lambda^{\pi}=(0^-,1^-,2^-)$ obtained by 
coupling a $1p_{\Xi^-}$ state to $J^{\pi}(^{14}$N$_{\rm g.s.})$=1$^+$, as 
shown in Fig.~\ref{fig:14N}. 

The energy splittings marked in Fig.~\ref{fig:14N} follow from a shell-model 
quadrupole-quadrupole spin independent residual interaction
${\cal V}_{\Xi N}$,
\begin{equation} 
{\cal V}_{\Xi N}=F^{(2)}_{\Xi N} Q_N\cdot Q_{\Xi}, \,\,\,\,\,\, 
Q_B=\sqrt{\frac{4\pi}{5}} Y_2({\hat{r}}_B), 
\label{eq:QdotQ}
\end{equation}
where $F^{(2)}$ is the corresponding Slater integral. A representative 
value of $F^{(2)}_{\Xi N}=-3$~MeV is used here, smaller than the value 
$F^{(2)}_{\Lambda N}=-3.7$~MeV established empirically for $p$-shell
$\Lambda$ hypernuclei~\cite{DalGal81}, in accordance with a $\Xi N$ strong
interaction somewhat weaker than the $\Lambda N$ strong interaction. 
A single $^3D_1$ $^{14}$N$_{\rm g.s.}$ component providing a good 
approximation to the full intermediate-coupling g.s. 
wavefunction~\cite{Mill07} was assumed in the present evaluation.

Fig.~\ref{fig:14N} shows a triplet of $^{14}$N$_{\rm g.s.}+1p_{\Xi^-}$ levels, 
spread over more than 1~MeV. The least bound of these states, $\Lambda^{\pi}=
0^-$, is shifted upward by 0.84~MeV from the $(2\Lambda+1)$ averaged position 
at $-1.96\pm 0.26$~MeV to $E(0^-)=-1.12\pm 0.26$~MeV. This is consistent 
with the averaged position $\bar{E}=-1.15\pm 0.20$~MeV of the three 
$\Xi^-\,^{14}$N$_{\rm g.s.}$ capture events listed in Table~\ref{tab:twin}. 
The $\Lambda^{\pi}=0^-$ state assumes spin-parity $J^{\pi}=\frac{1}{2}^-$ 
when Pauli-spin $s_{\Xi^-}=\frac{1}{2}$ is introduced, but its position is 
unaffected by spin dependent $\Xi N$ residual interactions in leading order. 
We are not aware of any good reason why capture has not been seen from the 
other two states with $\Lambda^{\pi}=1^-,2^-$. This may change when more 
events are collected at the next stage of the ongoing J-PARC E07 emulsion 
experiment.

\section{$1s_{\Xi^-}$ states in $^{14}$N?} 
\label{sec:KINKA} 

\begin{figure}[!h] 
\begin{center} 
\includegraphics[width=0.7\textwidth]{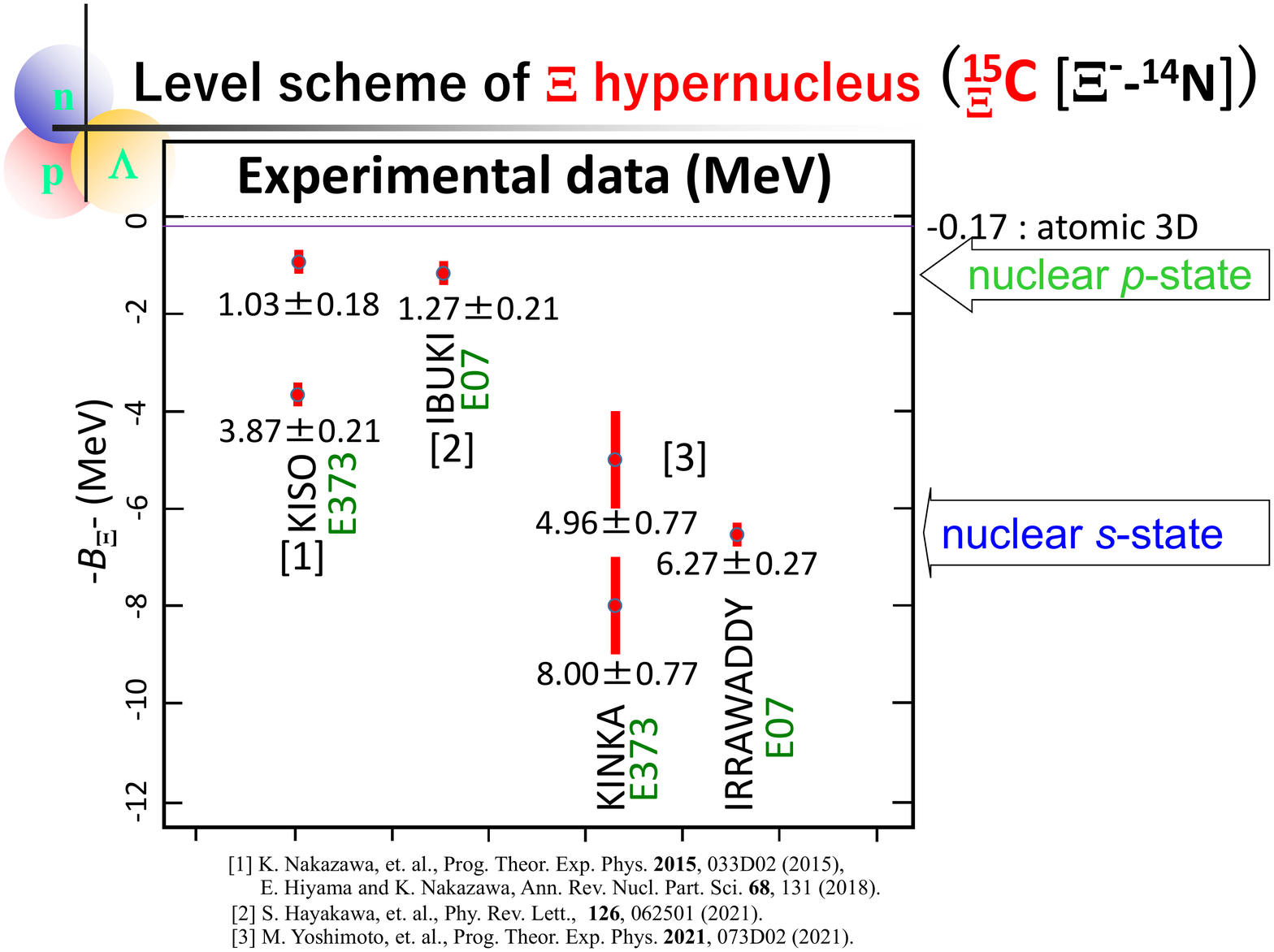} 
\caption{$\Xi^-_{1s}$ and $\Xi^-_{1p}$ nuclear states in $^{14}$N, assigned 
in KEK-E373 and J-PARC E07 emulsion experiments by interpreting $\Xi^-$ 
capture events that lead to observed twin-$\Lambda$ hypernuclear decays. 
Figure provided by Dr. K.~Nakazawa, based on recent results from 
Refs.~\cite{E373b,E07a,E07b}.} 
\label{fig:nakazawa} 
\end{center} 
\end{figure}

In addition to the $\Xi^-_{1p}$--$^{14}$N capture events listed as KISO and 
IBUKI in Table~\ref{tab:twin}, the J-PARC light-nuclei emulsion experiment 
E07 reported also two other events KINKA and IRRAWADDY, assigned as 
$\Xi^-_{1s}$--$^{14}$N states, see Fig.~\ref{fig:nakazawa}. We note that 
$2P\to 1S$ radiative decay rates are of order 1\% of $3D\to 2P$ radiative 
decay rates~\cite{Zhu91,Koike17} suggesting that $\Xi^-$ capture from 
a nuclear $\Xi^-_{1s}$--$^{14}$N state is suppressed to this order 
relative to capture from a nuclear $\Xi^-_{1p}$--$^{14}$N state. 
Assigning a $\Xi^-_{1s}$--$^{14}$N bound state to IRRAWADDY, and by default 
also KINKA which--given its large uncertainty--is not inconsistent with 
IRRAWADDY, is therefore questionable. 

It has been conjectured by us recently~\cite{FG23} that IRRAWADDY is a 
near-threshold $\Xi^0_{1p}$--$^{14}$C bound state that has nothing to do 
with a $\Xi^-_{1s}$--$^{14}$N bound state suggested by E07. The mixing 
induced by the $\Xi N$ strong interaction between a $\Xi^-_{1p}$--$^{14}$N
bound state identified with IBUKI and a $\Xi^0_{1p}$--$^{14}$C bound state 
lying about 5 MeV below IBUKU, within the J-PARC E07 experimental uncertainty 
of IRRAWADDY, is sufficiently strong to make the $E1$ radiative deexcitation 
of the $\Xi^-_{3D}$--$^{14}$N atomic state populate equally well {\it both} 
$\Xi^-_{1p}$--$^{14}$N and $\Xi^0_{1p}$--$^{14}$C nuclear states.

\section{Discussion}
\label{sec:disc}

\begin{table}[htb]
\begin{center}
\caption{$\Xi^-$--$^{12}$C and $\Xi^-$--$^{14}$N binding energies in 
$1s$ and $1p$ states, $B_{\Xi^-}^{1s}$ and $B_{\Xi^-}^{1p}$, plus $\Xi$ 
nuclear potential depths $V_{\Xi}(\rho_0)$ at nuclear-matter density 
$\rho_0=0.17\,$fm$^{-3}$ calculated using a density-dependent optical 
potential Eqs.~(\ref{eq:Vopt},\ref{eq:WRW}) with Re$\,b_0$ fitted to 
binding energies underlined for each input choice. All entries are in MeV.} 
\begin{tabular}{cccccc}
\hline
Input & $B_{\Xi^-}^{1s}(^{12}$C) & $B_{\Xi^-}^{1p}(^{12}$C) & 
$B_{\Xi^-}^{1s}(^{14}$N) & $B_{\Xi^-}^{1p}(^{14}$N) & $V_{\Xi}(\rho_0)$  \\
\hline
KEK E176~\cite{E176} & 9.82 & $\underline{0.82}$ & 11.78 & 1.96 & 21.9  \\
J-PARC E07~\cite{E07b} & 4.94 & 0.31 & $\underline{6.27}$ & 0.50 &
13.8  \\
\hline
\end{tabular}
\label{tab:targets}
\end{center}
\end{table}

Two $\Xi$-nuclear scenarios are listed in Table~\ref{tab:targets}. In the 
first one, two KEK-E176 $^{12}$C events~\cite{E176}, with $B_{\Xi^-}^{1p}=
0.82\pm 0.15$~MeV, serve as input for setting up the strength of the 
$\Xi$-nuclear optical potential. Other $\Xi^-$ binding energies are then 
predicted, as listed in the first row of Table~\ref{tab:targets}. We note that 
a value $V_{\Xi}\gtrsim 20$~MeV implies a substantially stronger in-medium 
$\Xi N$ attraction than reported by some recent free-space model evaluations 
(HAL-QCD~\cite{HALQCD19}, EFT@NLO~\cite{HM19,K19} and RMF~\cite{Gaitanos21}),
all of which satisfy $V_{\Xi}\lesssim 10$~MeV. A notable exception is provided
by versions ESC16*(A,B) of the latest Nijmegen extended-soft-core $\Xi N$
interaction model~\cite{ESC16}, in which values of $V_{\Xi}$ higher than
20~MeV are derived. However, these large values are reduced substantially
by $\Xi NN$ three-body contributions within the same ESC16* model.

Choosing instead the J-PARC E07 $^{14}$N IRRAWADDY event, with $B_{\Xi^-}^{1s}
=6.27\pm 0.27$~MeV~\cite{E07b} as input, gives rise to different predictions 
as listed in the second row of the table. The difference between the two 
sets of predictions is striking, particularly for the $\Xi^-_{1s}$ binding 
energies. This large difference is reflected also in the $\Xi$-nuclear 
potential depths at nuclear matter density, $V_{\Xi}(\rho_0)$, listed in the 
last column of the table. Equally interesting is the difference between the 
two sets with regard to $\Xi^-_{1p}$ bound states. In particular, the 
$\Xi^-_{1p}$--$^{12}$C binding energy constrained by $B_{\Xi^-}^{1s} 
(^{14}$N)=6.27$\pm$0.27~MeV comes out in the second row exceedingly small, 
substantially disagreeing with that determined from the two KEK E176 capture 
events~\cite{E176} underlined in the first row. As for the calculated 
$\Xi^-_{1p}$--$^{14}$N binding energies, since $J({^{14}{\rm N}})\neq 0$, 
see Fig.~\ref{fig:14N}, these cannot be compared {\it directly} with IBUKI's 
binding energy of 1.27$\pm$0.21~MeV from Fig.~\ref{fig:nakazawa}. 

Recent Skyrme-Hartree-Fock (SHF) calculations~\cite{GZS21} presented global 
fits to comprehensive $\Xi$-nuclear data, including $B_{\Xi^-}^{1s}$=8.00~MeV 
for KINKA and $B_{\Xi^-}^{1p}$=1.13~MeV for the mean of KISO and IBUKI. Apart 
from small nonlocal potential terms and effective mass corrections, the SHF 
$\Xi$-nuclear mean-field potential $V_{\Xi}(\rho_N)$ consists of two terms: 
$V_{\Xi}^{(2)}(\rho_N)\propto\rho_N$ \& $V_{\Xi}^{(3)}(\rho_N)\propto\rho^2_N
$. Large-scale SHF fits of the corresponding $\Xi$ potential depths, and their 
sum, are listed in the first row of Table~\ref{tab:SHF}. Listed in the lower 
rows are $\Xi$ potential depths obtained in the optical potential methodology 
\cite{FG22} when fitting just the two $\Xi^-$ states in $^{14}$N as used in 
the SHF calculations~\cite{GZS21}. Good agreement is observed between the two 
methods, with $V_{\Xi}(\rho_0)\approx\,$14~MeV, similar to the depth value 
listed in the second row of Table~\ref{tab:targets} using IRRAWADDY alone. 
We note that the $\Xi^-_{1p}$ state in $^{12}$C (see Table~\ref{tab:targets}) 
comes out unbound in the SHF calculations unless $^{12}$C is made artificially 
deformed~\cite{GZS21}. 

\begin{table}[htb]
\begin{center}
\caption{$\Xi$-nuclear potential depths (in MeV) from a large-scale SHF 
fit~\cite{GZS21} and from our $V_{\rm opt}$ two-parameter fits to just 
$B_{\Xi^-}^{1s}$=8.00~MeV (KINKA) and $B_{\Xi^-}^{1p}$=1.13~MeV (KISO,IBUKI) 
in $^{14}$N. `Pauli' refers to Eq.~(\ref{eq:WRW}).} 
\begin{tabular}{ccccc}
\hline 
Method & Pauli & $V_{\Xi}^{(2)}(\rho_0)$ & $V_{\Xi}^{(3)}(\rho_0)$ & 
$V_{\Xi}(\rho_0)$ \\
\hline
SHF & No & 34.1 & $-$20.4 & 13.7  \\
$V_{\rm opt}$ & No & 27.5 & $-$12.6 & 14.9  \\
$V_{\rm opt}$ & Yes & 24.6 & $-$11.0 & 13.6 \\ 
\hline
\end{tabular}
\label{tab:SHF}
\end{center}
\end{table}

The solution proposed here to the difficulty of interpreting IRRAWADDY as
a $\Xi^-_{1s}$ bound state in $^{14}$N is by pointing out that it could
correspond to a $\Xi^0_{1p}$--$^{14}$C bound state, something that cannot
occur kinematically in the other light-emulsion nuclei $^{12}$C and $^{16}$O. 
Given that in this nuclear mass range capture rates from $1s_{\Xi^-}$ states 
are estimated to be two orders of magnitude below capture rates from 
$1p_{\Xi^-}$ states~\cite{Zhu91,Koike17}, this $\Xi^0_{1p}$--$^{14}$C new 
assignment addresses satisfactorily the capture rate hierarchy. 
%Establishing $1s_{\Xi^-}$ states in light emulsion nuclei requires focusing 
%on capture events in $^{12}$C and $^{16}$O, where the $\Xi^-p\leftrightarrow
%\Xi^0n$ coupling is ineffective. 

\section*{Acknowledgments} 
\begin{acknowledgement} 
One of us (A.G.) thanks Ji\v{r}\'{i} Mare\v{s} and other members of the
HYP2022 organizing team for their generous hospitality during the Conference. 
This work is part of a project funded by the EU Horizon 2020 Research 
\& Innovation Programme under grant agreement 824093. 
\end{acknowledgement}

\end{document}